\documentclass[epj,nopacs]{svjour}
\usepackage{epsfig}
\usepackage{latexsym}
\usepackage[colorlinks=true,linktocpage=true,linkcolor=blue,citecolor=blue,allcolors=blue]{hyperref}
\usepackage{url}
\usepackage[utf8]{inputenc}
\usepackage{enumerate}
\usepackage{color}
\usepackage{xcolor}
\usepackage{bm}

\usepackage{amsmath}
\usepackage{amssymb}

\usepackage[english]{babel}
\usepackage{url}
\usepackage[normalem]{ulem}
\newcommand{\txt}[1]{\textrm{#1}}

\begin{document}

 \title{
 The high-density equation of state in heavy-ion collisions: \\
 Constraints from proton flow
 }

 \titlerunning{The Eos from proton flow}
 
\author{Jan~Steinheimer\inst{1} \and Anton Motornenko \inst{1} \and  Agnieszka Sorensen\inst{2} \and Yasushi~Nara\inst{3} \and Volker~Koch\inst{4} \and Marcus~Bleicher\inst{5,6,7}
}       

\institute{Frankfurt Institute for Advanced Studies, Ruth-Moufang-Str. 1, D-60438 Frankfurt am Main, Germany \and Institute for Nuclear Theory, University of Washington, Box 351550, Seattle, Washington 98195, USA \and Akita International University, Yuwa, Akita-city 010-1292, Japan \and Lawrence Berkeley National Laboratory, 1 Cyclotron Road, Berkeley, California 94720, USA \and Institut f\"{u}r Theoretische Physik, Goethe Universit\"{a}t Frankfurt, Max-von-Laue-Str. 1, D-60438 Frankfurt am Main, Germany \and GSI Helmholtzzentrum f\"ur Schwerionenforschung GmbH, Planckstr. 1, D-64291 Darmstadt, Germany \and Helmholtz Research Academy Hesse for FAIR (HFHF), GSI Helmholtzzentrum f\"ur Schwerionenforschung GmbH, Campus Frankfurt, Max-von-Laue-Str. 12, 60438 Frankfurt am Main, Germany}
 
%\author{Jan Steinheimer}
%\affiliation{Frankfurt Institute for Advanced Studies, Ruth-Moufang-Str. 1, D-60438 Frankfurt am Main, Germany}

%\author{Anton Motornenko}
%\affiliation{Frankfurt Institute for Advanced Studies, Ruth-Moufang-Str. 1, D-60438 Frankfurt am Main, Germany}

%\author{Agnieszka Sorensen}
%\affiliation{Institute for Nuclear Theory, University of Washington, Box 351550, Seattle, Washington 98195, USA}

%\author{Yasushi Nara}
%\affiliation{Akita International University, Yuwa, Akita-city 010-1292, Japan}

%\author{Volker Koch}
%\affiliation{%
%Lawrence Berkeley National Laboratory, 1 Cyclotron Road, Berkeley, %California 94720, USA
%}

%\author{Marcus Bleicher}
%\affiliation{Institut f\"{u}r Theoretische Physik, Goethe Universit\"{a}t Frankfurt, Max-von-Laue-Str. 1, D-60438 Frankfurt am Main, Germany}
%\affiliation{
%GSI Helmholtzzentrum f\"ur Schwerionenforschung GmbH, Planckstr. 1, D-64291 Darmstadt, Germany}
%\affiliation{
%Helmholtz Research Academy Hesse for FAIR (HFHF), GSI Helmholtzzentrum f\"ur Schwerionenforschung GmbH, Campus Frankfurt, Max-von-Laue-Str. 12, 60438 Frankfurt am Main, Germany}

\date{\today}

%\begin{abstract}

\abstract{A set of different equations of state is implemented in the molecular dynamics part of a non-equilibrium transport simulation (UrQMD) of heavy-ion collisions. It is shown how different flow observables are affected by the density dependence of the equation of state. In particular, the effects of a phase transition at high density are explored, including an expected reduction in mean $m_T$. We also show that an increase in $v_2$ is characteristic for a strong softening of the equation of state. The phase transitions with a low coexistence density, $n_{\txt{CE}}<4 n_0$, show a distinct minimum in the slope of the directed flow as a function of the beam energy, which would be a clear experimental signal. By comparing our results with experimental data, we can exclude any strong phase transition at densities below $4n_0$.
}
%\end{abstract}

\maketitle

\section{Introduction}

The role of the QCD equation of state (EoS) in collisions of heavy nuclei at various beam energies has been a central aspect of many years of dedicated research. While the EoS for vanishing net baryon density is now well constrained by lattice QCD calculations~\cite{Borsanyi:2013bia,HotQCD:2014kol,Bazavov:2017dus}, the high-density and intermediate- to low-temperature EoS is still not well known. 

Some information can be extracted from the properties of neutron stars, the densest objects known in the universe, as well as from binary collisions of these stars \cite{Tolman:1939jz,Oppenheimer:1939ne,Hinderer:2009ca,Bauswein:2013yna,Hanauske:2017oxo,Hanauske:2019qgs,HADES:2019auv,LIGO,Most:2018eaw}. To make the connection between these astrophysical observations, lattice QCD, and relativistic nuclear collisions is an important aspect of constructing a consistent and complete picture of the properties of strongly interacting matter.

At the highest beam energies available, the EoS of QCD can be studied in fluid dynamical approaches and it was shown that the dynamically-created flow in heavy-ion reactions is consistent with a smooth crossover as calculated by lattice QCD methods \cite{Pratt:2015zsa}. As we go to lower beam energies, these 'standard models of heavy-ion reactions' or hybrid models \cite{Paiva:1996nv,Hama:2004rr,Bass:2000ib,Hirano:2001eu,Kolb:2003dz,Hirano:2004en,Hirano:2012kj,Nonaka:2006yn,Petersen:2008dd,Werner:2010aa,Gale:2013da,Shen:2014vra}, incorporating both a fluid dynamic and a non-equilibrium transport model part, become less suitable for describing the system's evolution. This is because at lower energies, the approach to equilibration, necessary for the initial state used in the fluid dynamic approaches, takes an ever increasing fraction of the duration of the collision. Moreover, the role of the EoS during the compression phase is not addressed in hybrid models, and it has been shown to have a strong impact on the extracted observables \cite{Nara:2021fuu,OmanaKuttan:2022the}.

At the same time, direct first principle calculations of the QCD EoS at high density and intermediate to low temperatures are not possible due to the fermion sign problem \cite{deForcrand:2009zkb} and the fact that expansion schemes to finite density are reliable only to a certain degree \cite{Fodor:2002km,Allton:2002zi,Allton:2003vx,Allton:2005gk,Laermann:2003cv,DElia:2004at,DElia:2002gd,Borsanyi:2020fev,Borsanyi:2021sxv}.
Consequently, any discovery about the EoS must come from thorough comparisons of simulations of heavy-ion collisions as well as astrophysical observations, including an EoS that fulfills not only the constraints of lattice QCD at vanishing density, but also the observable constraints from heavy-ion reactions as well as neutron stars and their mergers.

Early, it was suggested that the flow, especially of baryons, may be sensitive to the EoS. In particular, a phase transition would leave characteristic imprints on the flow profile. For example, a strong phase transition would lead to a stalled radial expansion and thus a plateau in the mean transverse energy excitation function \cite{VanHove:1982vk}. Furthermore, the collapse of the flow, a distinct minimum in the the slope of the directed flow as a function of the beam energy, was predicted \cite{Stoecker:1986ci,Brachmann:1999xt,Ohnishi:2017xjg}. Later, also higher moments of the anisotropic flow were considered sensitive to the EoS \cite{Stoecker:1980vf,Danielewicz:2002pu,LeFevre:2016vpp,Nara:2017qcg,Nara:2018ijw}. 
Over the recent years, several studies within the standard hybrid model approaches showed that these signals were often washed out by the matching which is required between different parts of the models \cite{Steinheimer:2014pfa}. In particular, the transition between the equilibrium and non-equilibrium descriptions is especially questionable at low beam energies, where most of the signals are expected. 

Until now, only few attempts have been made to consistently introduce a phase transition (or any non-trivial features in the EoS) in a non-equilibrium simulation of heavy-ion collisions \cite{OmanaKuttan:2022the,Nara:2017qcg,Oliinychenko:2022uvy,Li:1998ze,Sorge:1998mk,Sorensen:2020ygf,Guo:2020mhy} at beam energies currently under investigation at the RHIC-BES and, in the future, at the upcoming SIS18/SIS100 accelerator. 

The purpose of the present work is to conduct a systematic study of the sensitivity of different flow observables to various EoSs, including EoSs with phase transitions, in a fully consistent microscopic transport approach. To do so, we will introduce a simple but consistent way to implement different EoSs, in particular some including a phase transition at high densities, in a microscopic transport model, and we will use this model to investigate the effects of such a phase transition on transverse flow observables. By comparing the simulation results for the different EoSs, we will be able to identify clear effects of a phase transition on the observables, while comparisons with available data may enable us to put better constraints on the existence of such a transition in the QCD EoS.

\section{Microscopic Transport with density-dependent potentials}
\label{ref:transport}

To systematically study the effects of a phase transition in nuclear collisions requires a model that contains a certain set of baseline features. This includes, for example, a proper treatment of the conservation laws and finite size effects as well as a treatment of non-equilibrium effects. In addition, the model should allow for the calculation of a multitude of hadronic observables in a consistent way.

In this work we will employ the non-equilibrium molecular dynamics (MD) implementation of the UrQMD transport approach~\cite{Bass:1998ca,Bleicher:1999xi}. It is based on the propagation of hadrons on classical trajectories in combination with stochastic binary scatterings, color string formation, and resonance excitation and decays. The imaginary part of hadron interactions are based on a geometric interpretation of their scattering cross sections, which are either taken from experimental measurements where available \cite{ParticleDataGroup:2020ssz}, or are calculated, e.g., from the principle of detailed balance\footnote{This means that we do not include any possible explicit momentum dependence of the MD potentials which would also require modifications of the effective elastic scattering cross sections at low relative momenta.}.
In its default setup, the model corresponds to a hadronic cascade and can be readily used to describe the final state spectra of hadrons over a wide range of beam energies. It was shown that the effective EoS of the UrQMD in cascade mode corresponds to a Hadron Resonance Gas (HRG) with the respective degrees of freedom \cite{Bravina:2008ra}.

Extending the equations of motion to include the real part of hadronic interactions is usually done with a quantum molecular dynamics (QMD) approach and a density-dependent potential interaction term. Unlike other mean-field approaches such as the Boltzmann--Uehling--Uhlenbeck (BUU) or Vlasov--Uehling--Uhlenbeck (VUU) transport, QMD is an $n$-body theory describing the interactions between $n$ nucleons. Early, non-relativistic QMD approaches~\cite{Aichelin:1991xy} were developed which incorporated density-dependent Skyrme interactions \cite{Hartnack:1997ez}.

In the QMD part of the UrQMD model, the change in the momenta of the baryons due to a density-dependent potential is calculated using the non-relativistic equations of motion,
\begin{eqnarray}\label{motion}
\dot{\textbf{r}}_{i}=\frac{\partial \mathrm{\bf{H}}  }{\partial\textbf{p}_{i}},
\quad \dot{\textbf{p}}_{i}=-\frac{\partial \mathrm{\bf{H}} }{\partial \textbf{r}_{i}},
\end{eqnarray}
where $ \mathrm{\bf{H}} = \sum_i H_i$ is the total Hamiltonian function of the system which is just the sum over all Hamiltonians, $H_i=E^{\mathrm{kin}}_i + V_i$, of the $i$ baryons.
It includes the kinetic energy and the total potential energy
$ {\mathrm{\bf{V}}=\sum_i V_i \equiv \sum_i V\big(n_B(r_i)\big)}$. The change of momentum of each baryon can be calculated from Hamilton's equations of motion,
\begin{eqnarray}
\dot{\textbf{p}}_{i} & = &-\frac{\partial \mathrm{\bf{H}}}{\partial \textbf{r}_{i}} =  -  \frac{ \partial \mathrm{\bf{V}} }{\partial \textbf{r}_{i}} \\
  & = & - \left(\frac{ \partial V_i }{\partial n_i}\cdot \frac{\partial n_i}{\partial \textbf{r}_{i}} \right)-\left( \sum_{j \ne i} \frac{\partial V_j}{ \partial n_j} \cdot \frac{\partial n_j}{\partial \textbf{r}_{i}}\right) ~,
\label{QMD_eq}
\end{eqnarray}
where $n_{\{i,j\}} \equiv n_B (\bm{r}_{\{i,j\}})$ is the local interaction density of baryon $i$ or $j$.
Thus, $V_i$ corresponds to the average potential energy of a baryon at position $\bm{r}_i$, and the local interaction density $n_B$ at position $\bm{r}_k$ is calculated by assuming that each particle can be treated as a Gaussian wave packet~\cite{Aichelin:1991xy,Bass:1998ca}. With such an assumption, the local interaction baryon density $n_B(\bm{r}_k)$ at location ${\bf r}_k$ of the $k$-th particle in the computational frame is:
\begin{eqnarray}
    n_B(\bm{r}_k) &=& n_k = \sum_{j,\,j\neq k} n_{j,k} \\ \nonumber 
    & = & \left(\frac{\alpha}{\pi}\right)^{3/2}\sum_{j,\,j\neq k} B_j \exp{\left(-\alpha({\bf r}_k-{\bf r}_j)^2\right)} \, , \label{gauss_qmd}
\end{eqnarray}
where $\alpha=\frac{1}{2 L}$, with $L=2$ fm$^2$, is the effective range parameter of the interaction. The summation runs over all baryons, and $B_j$ is the baryon charge of the $j$-th baryon. Once the potential energy per baryon is know equation (\ref{QMD_eq}) can be solved numerically.

The density dependent average potential energy per baryon, $V$, can be related to a density-dependent single-particle potential $U$, which is sometimes given instead, through
\begin{equation}
\label{eq:U}
U(n_B)=\frac{\partial \big(n_B  \cdot V(n_B)\big)}{\partial n_B}\,.
\end{equation}
For a given particle, the single-particle potential is then also given by $U_i = U \big( n_B(\bm{r}_i)\big)$.

Note that the above implies that the force that the $i$-th baryon is subjected to depends not only on the change of the potential energy at point $\bm{r}_i$ due to the local gradient of $n_B(\bm{r}_i)$, but also on the change of the potential at positions $\bm{r}_j$ of all baryons $j$ due to the change in $\bm{r}_i$. The derivative of the potential in Eq.\ (\ref{QMD_eq}), and thus $\dot{\textbf{p}}_{i}$, is calculated by summing over all possible baryon pairs $j \neq i$; note that by construction, we have $\sum_i \dot{\bm{p}}_i = 0$, that is momentum is conserved. 

In the following, the QMD implementation will assume, for simplicity, that the mean-field potential for all baryon types is the same as that for the nucleons. 

In the present implementation of the QMD model we ignore relativistic effects, which should only become relevant at beam energies between $\sqrt{s_{\mathrm{NN}}} = 5$ -- 10 GeV~\cite{Nara:2019qfd}.

\subsection{Equations of state}
 
The relevant input for the QMD equations of motion is the density-dependent interaction potential $V(n_B)$ which determines the effective EoS governing the evolution. In this paper, we will compare EoSs coming from a variety of models: the Skyrme model, the vector density functional (VDF) model, as well as a realistic Chiral Mean Field~(CMF) model which incorporates all interactions essential for a realistic description of nuclear matter, neutron stars, and hot QCD matter.
  
\subsubsection{The Skyrme model} 
 
In the original Skyrme UrQMD approach \cite{Aichelin:1991xy,Bass:1998ca} the density dependence of the single-particle energy for all baryons is given by a simple form:
\begin{equation}
\label{eq:usky}
    U_{\mathrm{Skyrme}}(n_B)= \alpha (n_B/n_0) + \beta (n_B/n_0)^{\gamma}\,.
\end{equation}
Two of the degrees of freedom that are provided by the three parameters ($\alpha$, $\beta$ and $\gamma$) can be constrained by the nuclear matter saturation density and binding energy, while the last remaining degree of freedom is constrained by the nuclear incompressibility, often referred to as the stiffness of the EoS. To obtain the average potential energy $V_{\mathrm{Skyrme}}(n_B)$ for the equations of motion, equation (\ref{eq:U}) can be integrated assuming that the potential energy $V_{\mathrm{Skyrme}}(n_B=0) =0$ vanishes at zero baryon density. As a baseline for comparisons, we will use two different Skyrme parametrizations denoted as \textit{hard} and \textit{soft} Skyrme EoS, with the nuclear incompressibility $K_0=380$ MeV and $200$ MeV, respectively. The parameters corresponding to the \textit{hard} and \textit{soft} parametrizations are shown in appendix \ref{s-para}.
The Skyrme EoS for densities above the nuclear saturation density is therefore fixed by parameters which are defined at the saturation density. Despite its simplicity, the Skyrme EoS has been used widely in many efforts to constrain the EoS close to saturation density. Its main shortcoming is the fact that by fixing the incompressibility, the high density behavior is also uniquely determined. Since presently rather precise constraints on the incompressibility are known, the Skyrme potential is essentially without any degree of freedom. To be able to study the EoS for densities far above nuclear saturation, it would be beneficial to extend the potential beyond a simple two-term Skyrme prescription (see also, e.g., \cite{Nara:2019qfd,Nara:2020ztb,OmanaKuttan:2022the} for a discussion and another possible way to extend the EoS).

\subsubsection{The vector density functional model (VDF)}
 
Recently, a vector density functional (VDF) model of the EoS has been developed \cite{Sorensen:2020ygf} which generalizes Eq.\ (\ref{eq:usky}) in two ways. First, the VDF model is relativistically covariant and therefore leads to fully covariant equations of motion. Second, the VDF model allows one to include an arbitrary number of interaction terms, which in turn enables description of non-trivial features such as, e.g., a phase transition at high baryon density and high temperature, while at the same time satisfying known constraints at low temperatures and low densities, e.g., reproducing the values of the nuclear saturation density $n_0$, binding energy at saturation $E_0$, and the location of the critical point of the nuclear liquid-gas phase transition $(n_c^{(N)}, T_c^{(N)})$.

\begin{table*}[t]
\centering
\begin{tabular}{|c||c|c|c|c|c||c|c|c|}
\hline
\hline
EoS  & $T_{c}^{(N)} [\txt{MeV}]$ &  ~$n_{c}^{(Q)} [n_0]$~ & $T_{c}^{(Q)} [\txt{MeV}]$ & ~~$\eta_L [n_0]$~~ & ~~$\eta_R [n_0]$ ~~&  $K_0 [\txt{MeV}]$ & ~~$n_L[n_0]$~~ & ~~$n_R[n_0]$~~ \\
\hline
 VDF1  &  18 & 3.0 & 100 & 2.50 & 3.315 & 261 & 2.13 & 3.57 \\
 \hline
 VDF2  & 18 &4.0 & 50 & 3.85 & 4.124 & 279 & 3.74 & 4.22 \\
 \hline
 VDF3  &  22 &6.0 & 50 & 5.80 &  6.177 & 356 & 5.66 & 6.31 \\
\hline
\hline
\end{tabular}
\caption{Characteristics of the VDF EoSs used in this work. For all VDF EoSs, the saturation density, the binding energy, and the critical density of the nuclear liquid-gas phase transition have been set at $n_0 = 0.160 ~ \mathrm{fm}^{-3}$, $E_0 = - 16.3 ~ \mathrm{MeV}$, and $n_c^{(N)} = 0.375n_0$, respectively. 
The incompressibility $K_0$ and the boundaries of the coexistence region of the ``QGP-like'' phase transition at $T=0$, denoted with $n_L$ and $n_R$, are not used to fit the parameters, but rather are properties of the EoSs resulting from fitting to the chosen set of characteristics $\{n_0, E_0, n_{c}^{(N)}, ~ T_{c}^{(N)}, ~ n_{c}^{(Q)}, ~ T_{c}^{(Q)}, ~ \eta_L, ~ \eta_R \}$. Parameters of the VDF model leading to the above sets of characteristics are listed in Appendix \ref{VDF_parameter_sets}.
}
\label{t1_VDF}
\end{table*}
 
The VDF energy density of a system composed of one species of baryons with degeneracy $g$, interacting through a number $K$ of mean-field vector interaction terms, is given by
\begin{eqnarray}
\mathcal{E}_{K} = \mathcal{E}_{\txt{kin}} + \sum_{k = 1}^K \left[ A_k^0 j_0 - g^{00} \left(\frac{b_k - 1}{b_k} \right) A_k^{\mu} j_{\mu}   \right]  
\end{eqnarray}
where the kinetic contribution to the total energy of the system is given by
\begin{eqnarray}
\mathcal{E}_{\txt{kin}} =  g\int \frac{d^3 p}{(2\pi)^3} ~  \epsilon_{\txt{kin}} ~ f_{\bm{p}} ~,
\end{eqnarray}
with $f_{\bm{p}}$ denoting the distribution function of the system, while the vector fields are defined as
\begin{eqnarray}
A_k^{\mu} \equiv C_k \big( j_{\nu} j^{\nu} \big)^{\frac{b_k}{2} - 1} j^{\mu}~,
\end{eqnarray}
where $j^{\mu}$ is the conserved current and $\big( j_{\nu} j^{\nu} \big)^\frac{1}{2} = n_B$ is the rest frame baryon density and $C_k$ are the parameters of the model. It is important to note that the quasiparticle kinetic energy is influenced by the presence of the vector fields, 
\begin{eqnarray}
\epsilon_{\txt{kin}} = \sqrt{ \left( \bm{p} - \sum_{k=1}^K \bm{A}_k \right)^2  + m^2  }~,
\end{eqnarray}
and similarly the spatial components of the conserved current are given by
\begin{eqnarray}
\bm{j} = g\int \frac{d^3 p}{(2\pi)^3} ~ \frac{\bm{p} - \sum_{k=1}^K \bm{A}_k}{\epsilon_{\txt{kin}}} ~ f_{\bm{p}} ~. 
\end{eqnarray}

From this, it is straightforward to calculate the quasiparticle energy,
\begin{eqnarray}
\varepsilon_{\bm{p}} \equiv \frac{\delta \mathcal{E}_K }{\delta f_{\bm{p}}} = \epsilon_{\txt{kin}} + \sum_{k=1}^K A_k^0 ~, 
\end{eqnarray}
which can be then taken as the quasiparticle Hamiltonian, leading to fully relativistic quasiparticle equations of motions (for details, see \cite{Sorensen:2020ygf}). The parameters of the model $\big\{C_k, b_k \big\}$ can be fixed by fitting the pressure of the system,
\begin{eqnarray}
 && P_K = g \int \frac{d^3p}{(2\pi)^3} ~ T ~ \ln\bigg[1 + e^{-\beta \big(\varepsilon_{\bm{p}} - \mu\big)} \bigg] \nonumber \\
 && \hspace{15mm} + ~\sum_{k=1}^K C_k \left( \frac{b_k - 1}{b_k} \right) n_{\scriptscriptstyle B}^{b_k}\,,
\end{eqnarray}
 to satisfy a chosen set of $2K$ constraints that establish the properties (characteristics) of nuclear matter, such as the saturation density or the location of the critical point.
 
 In the minimal version of the VDF model allowing one to describe nuclear matter with two first-order phase transitions (one corresponding to the ordinary nuclear liquid-gas phase transition, and the other corresponding to a postulated ``QGP-like'' phase transition at high density and temperature), the number of interaction terms is set at $K=4$. Then the parameters of the interactions are fit to reproduce the chosen values of the saturation density $n_0$, binding energy $E_0$, the critical density $n_c^{(N)}$ and temperature $T_c^{(N)}$ of the nuclear liquid-gas phase transition, the critical density $n_c^{(Q)}$ and temperature $T_c^{(Q)}$ of the postulated ``QGP-like'' phase transition, and the $T=0$ boundaries of the spinodal region of the postulated ``QGP-like'' phase transition, $\eta_L$ and $\eta_R$. To a large extent, the values of these characteristics of nuclear matter can be freely varied in the VDF model, although in practice there are some subtleties (see \cite{Sorensen:2020ygf} for more details). 
 
In this study, we used three different parametrizations of the VDF model, corresponding to three sets of characteristics of nuclear matter. For all sets, we took the characteristics of the ordinary nuclear matter to be $n_0 = 0.160 \ \txt{fm}^{-3}$, $E_0 = -16.3 \ \txt{MeV}$, and $n_c^{(N)} = 0.375n_0$. We list the remaining characteristics, varied among the different parametrizations, along with the resulting incompressibilities $K_0$, in Table \ref{t1_VDF}. We note here that in particular, as can be shown in a simple model of the nuclear matter EoS \cite{Kapusta:1984ij}, varying the critical temperature of the nuclear liquid-gas phase transition directly influences the resulting incompressibility $K_0$; this effect can be clearly seen in VDF3. Additionally, the high-density characteristics for VDF2 and VDF3, in particular the relatively small critical temperatures $T_c^{(Q)}$ and relatively narrow spinodal regions $(\eta_L, \eta_R)$, have been chosen to maximize the incompressibility of the EoS at intermediate densities. Parameters of the VDF model leading to the sets of characteristics given in Table \ref{t1_VDF} are listed in Appendix \ref{VDF_parameter_sets}.
 
Finally, while the VDF model has been designed for use in a BUU mean-field transport and as such is implemented in the hadronic transport code \texttt{SMASH} \cite{Weil:2016zrk} (publicly available at \cite{SMASH_website}), here it is used according to the paradigm explained above. In particular, we take
\begin{equation}
V_{\txt{VDF}} \equiv \frac{ \mathcal{E}_{\txt{MF}}\big|_{\substack{\txt{rest}\\\txt{frame}}} }{n_B}~,
\end{equation}
where $\mathcal{E}_{\txt{MF}}\big|_{\substack{\txt{rest}\\\txt{frame}}} = \sum_i \frac{C_i}{b_i} n_B^{b_i}$ is the total mean-field energy in the rest frame.

\subsubsection{The CMF model} 
The Chiral Mean Field model~\cite{Papazoglou:1998vr,Steinheimer:2010ib,Motornenko:2019arp} (CMF) is an approach to the description of QCD thermodynamics for a wide range of temperatures and densities. The effective degrees of freedom of the CMF model include a complete list of all known hadrons as well as the three light quark flavors plus a gluon contribution. The CMF contains a transition between quarks and hadronic degrees of freedom, the liquid-vapor transition in nuclear matter, as well as chiral symmetry restoration introduced through parity doubling in the mean field approximation. Parity doubling introduces the heavy parity partners to baryons of the lowest octet~\cite{Steinheimer:2011ea,Aarts:2018glk}. The baryons and their parity partners interact via mesonic mean fields (attractive scalar $\sigma,~\zeta$ and repulsive $\omega,~\rho,~\phi$ meson exchanges) and the effective masses of the parity partners become degenerate as chiral symmetry is restored. A detailed description of the CMF model and its parameters can be found in \cite{Motornenko:2020yme}.

The CMF model describes many aspects of QCD phenomenology. It has been successfully applied in an analysis of lattice QCD data~\cite{Motornenko:2020yme}, used in the description of cold neutron stars~\cite{Motornenko:2019arp}, and has been employed as the EoS in hydrodynamic simulations of both heavy-ion collisions and binary neutron star mergers~\cite{Seck:2020qbx,Most:2022wgo}.
 
The effective masses of the ground-state octet baryons and their parity partners (assuming isospin symmetry) read~\cite{Steinheimer:2011ea}:
\begin{eqnarray}
m^*_{b\pm} &=& \sqrt{ \left[ (g^{(1)}_{\sigma b} \sigma + g^{(1)}_{\zeta b}  \zeta )^2 + (m_0+n_s m_s)^2 \right]} \nonumber \\ 
  & \pm & g^{(2)}_{\sigma b} \sigma \ ,
\label{eq:mass}
\end{eqnarray}
where the various coupling constants $g^{(*)}_{*b}$ are determined by vacuum masses and by nuclear matter properties. $m_0$ refers to a bare mass term of the baryons which is not generated by the breaking of chiral symmetry, and $n_s m_s$ is the ${\rm SU}(3)_f$-breaking mass term that generates an explicit mass corresponding to the strangeness $n_s$ of the baryon. The single-particle energy of the baryons, therefore, becomes a function of their momentum $k$ and effective masses: $E^*=\sqrt{k^2+m_b^{* 2}}$.

Similar to the effective mass $m_b*$ which is modified by the scalar interactions, the vector interactions lead to a modification of the effective chemical potentials for the baryons and their parity partners:
\begin{equation}
      \mu^*_b=\mu_b-g_{\omega b} \omega-g_{\phi b} \phi-g_{\rho b} \rho\,.
      \label{eq:chem}
\end{equation}
Note that the couplings of nucleons and hyperons to the mean fields were fixed to reproduce a nuclear binding energy per baryon $E_0 = E/n_B - m_N^\mathrm{vac}$,
where $E$ is the energy density, as well as the asymmetry energy $S_0\approx31.9$~MeV and incompressibility $K_0\approx267$~MeV.

In the CMF model the single nucleon potential is given by the interactions with the chiral and repulsive mean fields. At $T=0$, it can be calculated from the self energy of the nucleons as:
\begin{equation}
\label{eq:cmfnp}
     \mathrm{U}_{\txt{CMF}}= m_{N}^{*} - m_{N}^{\txt{vac}} - \mu_{N}^{*} + \mu_{N} \,,
\end{equation} 
where $m_{N}^{\mathrm{vac}}$ and $\mu_{N}$ are the vacuum mass and chemical potential of the nucleon calculated only from the charge constraints and $m_{N}^{*}$ and $\mu_{N}^{*}$ are the corresponding effective nucleon mass, Eq.\ (\ref{eq:mass}), and effective chemical potential, Eq.\ \ref{eq:chem}), generated through the interactions with the scalar and vector mean fields.
 
In its default setup (referred to simply as CMF), the model exhibits a smooth transition from interacting hadronic to quark degrees of freedom. The speed of sound at $T=0$ shows a distinct maximum around 4-5 times saturation density due to the repulsive nuclear forces and then a slow softening followed by approaching the ideal gas limit from below \cite{Motornenko:2019arp}. 

Since the parameters of the CMF EoS are already constrained by lattice QCD data and neutron star observations, introducing an additional first-order phase transition consistently within the current framework of the model is challenging. Nevertheless, a simple augmentation can be used to implement a phase transition in the CMF EoS. To provide for another metastable state in the mean-field energy per baryon at large densities (in addition to the bound state present in ordinary nuclear matter below the critical point of the liquid-gas transition), the original potential of the CMF model is cut at density $n_B^{\txt{cut}}=$2.1$n_0$ for PT1 and 2.6$n_0$ for PT2, where PT1 and PT2 stand for two different phase transition scenarios. We then define the potential $\tilde{V}$ above $n_B = n_B^{\txt{cut}} + \Delta n_B$, where $\Delta n_B = 2.0n_0$ and $3.0n_0$ for PT1 and PT2, respectively, as $\tilde{V} (n_B) = V (n_B - \Delta n_B)$, effectively ``shifting'' the functional behavior of $V(n_B)$ by $\Delta n_B$. In the remaining gap between $n_B^{\txt{cut}}<n_B<n_B^{\txt{cut}}+\Delta n_B$, the mean-field energy is interpolated by a third order polynomial in order to create a second minimum in energy per particle $V(n_B)$ and at the same time ensure that its derivative is
a continuous function. In principle there are infinite ways to create such an construction, each leading to different properties of such constructed transition. Other methods can be employed to construct high-density phase transitions, for example using a vector density functional~\cite{Sorensen:2020ygf}. 

The present work is aimed at studying general signatures of a phase transition. The VDF EoSs all contain a critical endpoint at a moderate temperature. In contrast, for the CMF phase transition scenarios we choose an interpolation which results in a transition which is stronger than in the VDF model and in particular does not end in a critical endpoint at finite temperature. Here we use the fact that the phase structure at finite temperature is determined solely by the sum of the free hadron gas pressure and the pressure contribution of the fields, which is temperature-independent by construction.

As the phase transition is very strong, i.e., a too large drop in the pressure is created, the hadron gas pressure is never able to compensate for this drop and the phase transition remains present for all temperatures.
This means that, as we demand that the drop in the pressure for $n_B^{\txt{cut}} < n_B < n_B^{\txt{cut}} + \Delta n_B$ is large enough, due to our omission of an explicit temperature dependence, PT1 and PT2 will have no critical endpoint of the transition and remain a first-order transition for all collision systems. 

Since our procedure modifies the CMF EoS only at high densities, it leaves the low-density description consistent with the nuclear matter properties and lattice-QCD constraints.

\begin{figure}[t]
  \includegraphics[width=0.5\textwidth]{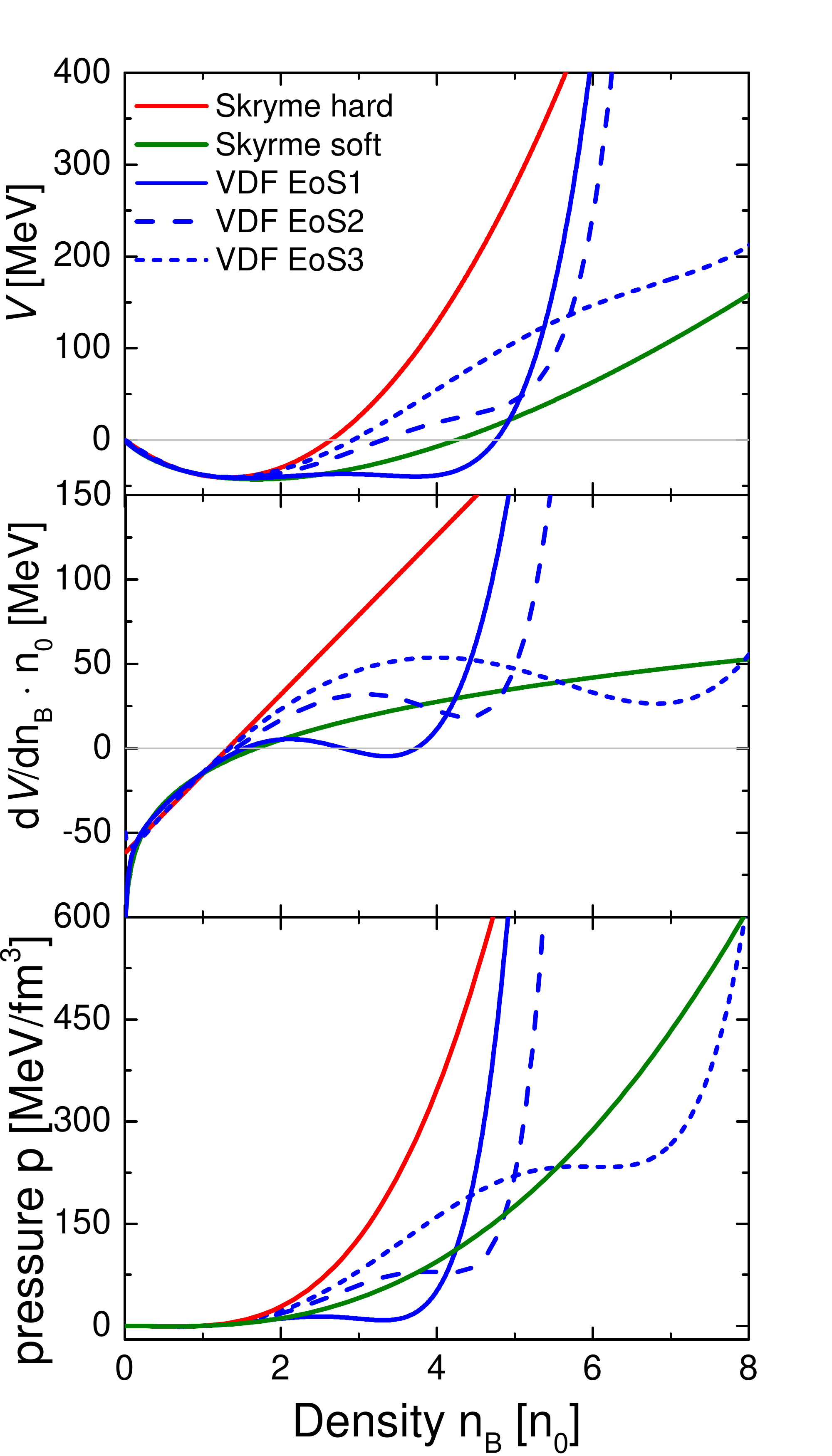}
  \caption{Upper plot: Field energy per baryon of three versions of the VDF EoS compared to those from the hard and soft Skyrme parametrizations. Middle plot: derivative of the field energy per baryon with respect to the baryon density, entering the calculation of the MD-forces, for the same EoSs. All VDF EoSs incorporate a phase transition (see Table \ref{t1_VDF}) and show significant unstable regions at different densities. Lower plot: pressure as function of the baryon density for all VDF EoSs, compared to the hard and soft Skyrme models.}
  \label{fig:VDF_eos}
\end{figure}

\begin{figure}[t]
  \includegraphics[width=0.5\textwidth]{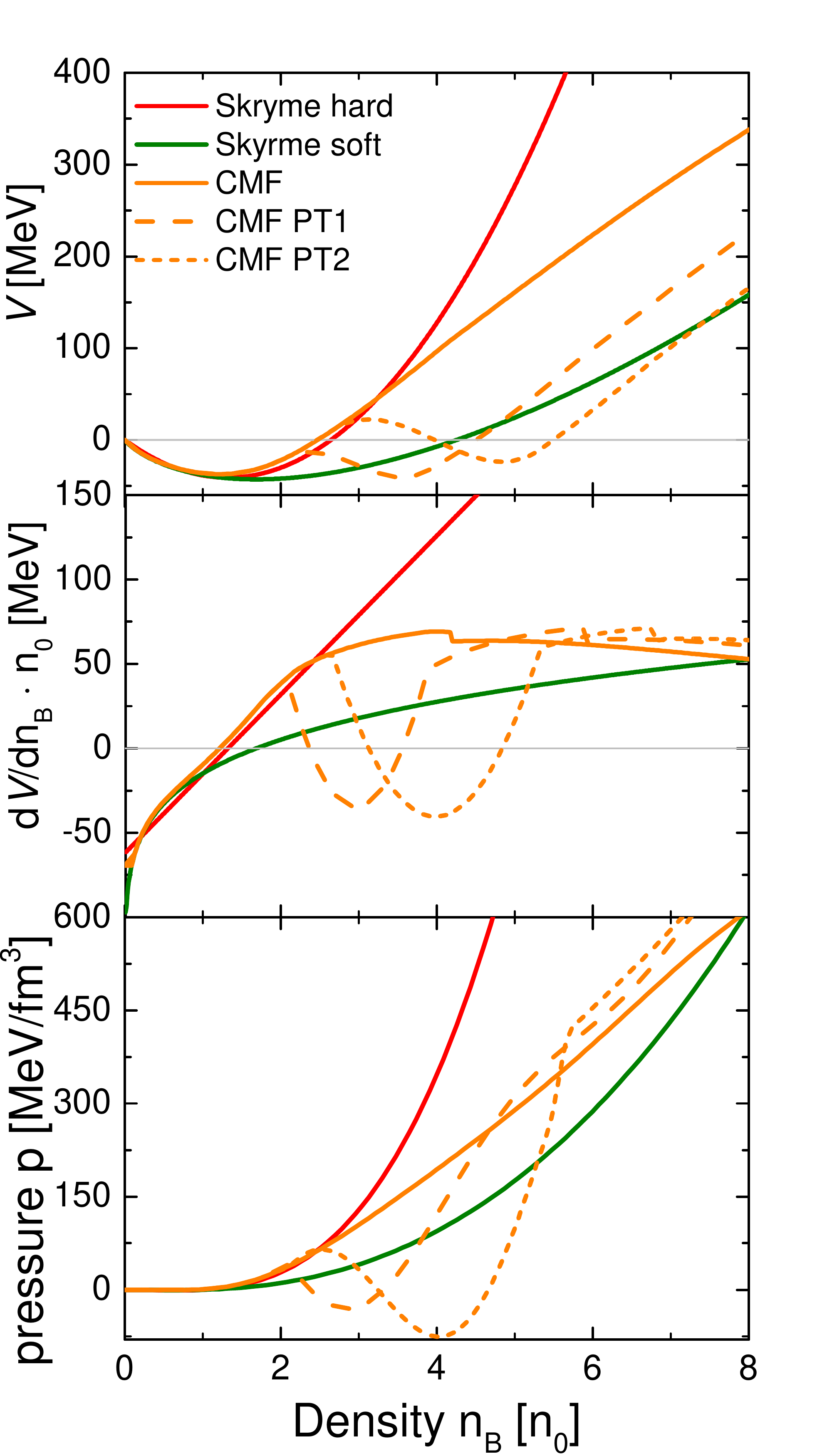}
  \caption{Upper plot: Field energy per baryon of three versions of the CMF EoS compared to those from the hard and soft Skyrme parametrizations. Middle plot: derivative of the field energy per baryon with respect to the baryon density, entering the calculation of the MD-forces, for the same EoSs. While the standard CMF essentially corresponds to a smooth transition with only a very minor phase transition at $\rho_B\approx 4 \rho_0$, the two scenarios PT1 and PT2 show significant unstable regions at different densities. Lower plot: pressure as a function of the baryon density for all CMF equations of state, compared to the hard and soft Skyrme models. 
  }
  \label{fig:cmf_eos}
\end{figure}

\begin{figure*}[t]
  \centering
  \includegraphics[width=\textwidth]{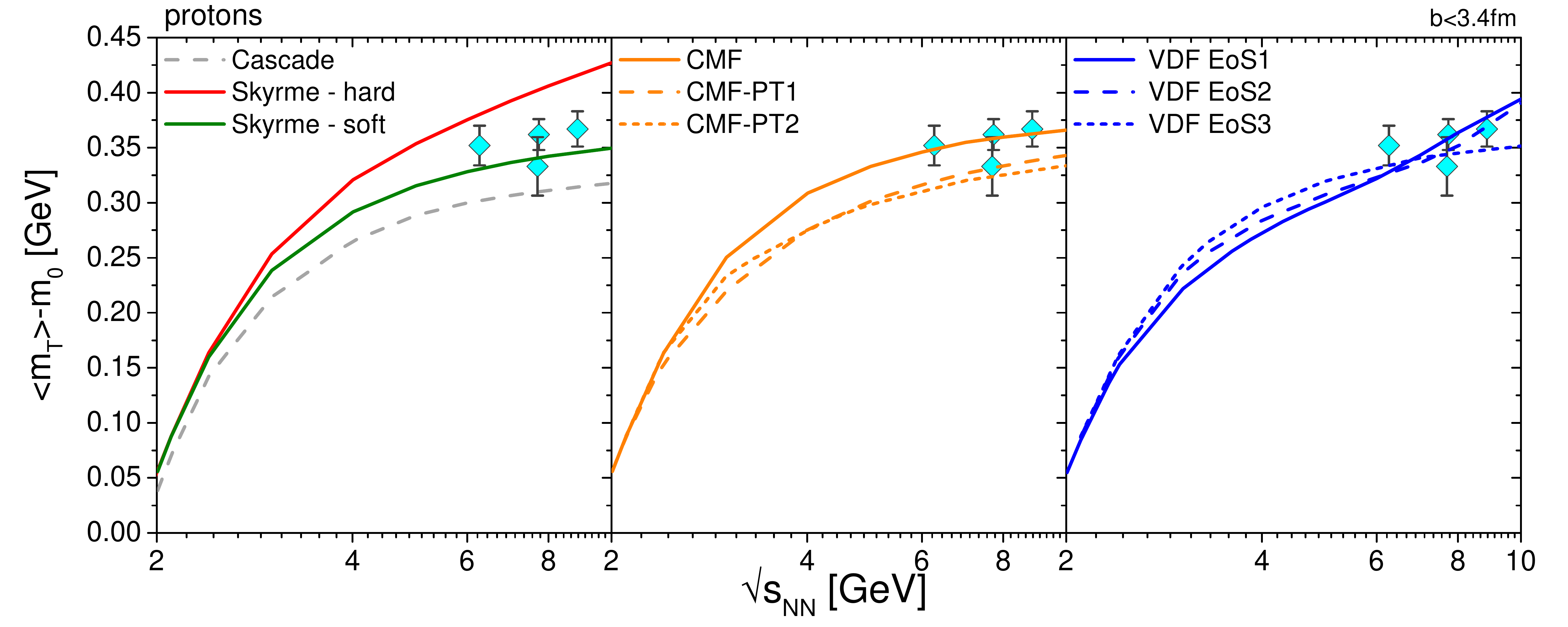}
  \caption{The mean transverse kinetic energy $\left\langle m_T \right \rangle - m_0$ of protons for central Au--Au collisions. $\left\langle m_T \right \rangle$ is evaluated at mid-rapidity $|y_\mathrm{cm}/y_b|<0.1 $ ($y$ is the proton rapidity in the center of mass frame of the collision, i.e. $y_\mathrm{cm} = 0$, and $y_b$ is the beam rapidity defined in the center-of-mass frame of the collision) and no cut in the transverse momentum is applied. From left to right, the results for the \texttt{UrQMD} cascade mode (grey dashed line), hard (red) and soft (green) Skyrme EoS, the CMF model (orange lines,) and the VDF model (blue lines) are shown. All are compared to available data from heavy-ion collisions at several beam energies (symbols) \cite{NA49:2006gaj,STAR:2017sal}.
  }
  \label{fig:mmt}
\end{figure*}

\subsubsection{Practical implementation of the EoS in \texttt{UrQMD}}

With the three above described models, we now have a wide range of possible EoSs at our disposal which either show no transition, a smooth crossover, or a phase transition with or without a critical endpoint.

To implement the above EoSs in the QMD part of the UrQMD model, we need to calculate the density dependence of the average field energy per baryon $V(n_B)$ within each model, which then can be used in the QMD equations of motion given by Eq.\ \eqref{QMD_eq}. In particular, $V(n_B)$ and its derivative need to be provided in order to numerically calculate changes in momentum at a given time-step.

The VDF model can directly provide the mean-field energy per baryon $V(n_B)$. In the CMF model, the nucleon interaction is described relativistically via scalar and vector mean fields which are not present in \texttt{UrQMD}. In addition, the CMF model is not only restricted to nucleons, thus, the single nucleon potential $\mathrm{U}_{\txt{CMF}}$ as defined in Eq.\ (\ref{eq:cmfnp}) is not suitable to calculate the relevant mean-field potential required for the equations of motion.
Fortunately, the effective field energy per baryon $E_{\mathrm{field}}/A$ can be used, i.e., the relevant quantity which enters the equations of motion is then defined as
\begin{equation}
    V_{\txt{CMF}} = E_{\mathrm{field}}/A  \equiv E_{\mathrm{CMF}}/A - E_{\mathrm{FFG}}/A\,,
\end{equation}
where $E_{\mathrm{CMF}}/A$ is the total energy per baryon at $T=0$ from the CMF model and $E_{\mathrm{FFG}}/A$ is the energy per baryon in a free non-interacting Fermi gas.
The resulting average field energy per baryon as a function of the baryon density, from both the VDF and CMF models, is shown in Figs.\ \ref{fig:VDF_eos} and \ref{fig:cmf_eos}, respectively. Here, the upper panels show the field energy per baryon $V$ as a function of the baryon density. In both cases, the EoSs are compared to the standard ``hard'' and ``soft'' Skyrme EoSs often used in the literature \cite{Hartnack:1997ez}. Due to the presence of a first-order transition at high densities, the VDF model shows generally a softer EoS than the hard Skyrme for $n_B \gtrsim n_0$. The standard CMF EoS shows a behavior similar to that of the soft Skyrme potential for sub-saturation (up to saturation) densities and a very stiff behavior at super-saturation densities up to ~2.5 $n_0$, followed by a softening. The two phase transitions included in PT1 and PT2 occur at different densities, as can be seen from the strong softening of $V$. It should be noted that these phase transitions can be considered very strong and therefore do not end in a critical endpoint as in the VDF model.

The middle panels of Figs.\ \ref{fig:VDF_eos} and \ref{fig:cmf_eos} show the derivative of the field energy per nucleon with respect to the baryon density as a function of baryon density $n_B$ in units of the ground-state baryon density for the different potentials. Again, the CMF PT1 nd PT2 clearly show the strongest minimum, and consequently one xpects stronger effects than with the VDF EoS. 

The actual equation of state, i.e., pressure as a function of the baryon density, is shown for $T=0$ in the lower panels of Figs.\ \ref{fig:VDF_eos} and \ref{fig:cmf_eos}. For all EoSs used in this study, the pressure can be obtained from
\begin{eqnarray}
 P(n_B) & = & P_{\rm id}(n_B) + \int_0^{n_B} n' \frac{\partial U(n')}{\partial n'} dn'\,.
\end{eqnarray}
where $ P_{\rm id}(n_B)$ is the pressure of an ideal Fermi gas of baryons. The density regions where the pressure decreases with density can be easily identified as the spinodal regions of the phase transitions.

Using the obtained potentials $V$ and their derivatives, it is now straightforward to solve the equations of motion Eq.\ \eqref{QMD_eq} in discrete time steps of $\Delta t= 0.2 \mathrm{fm/c}$. As described above, the baryons are treated as single Gaussian wavepackages centered around their positions $\bm{r}_i$. The time-evolution is always calculated in the center-of-velocity frame of the two incoming nuclei.

Having now established a method in which many characteristically different EoSs can be easily introduced in the QMD part of \texttt{UrQMD}, we will study the flow in heavy-ion collisions at beam energies from $\sqrt{s_{\mathrm{NN}}}= 2$--$10$ GeV.

\begin{figure*}[t]
  \centering
  \includegraphics[width=\textwidth]{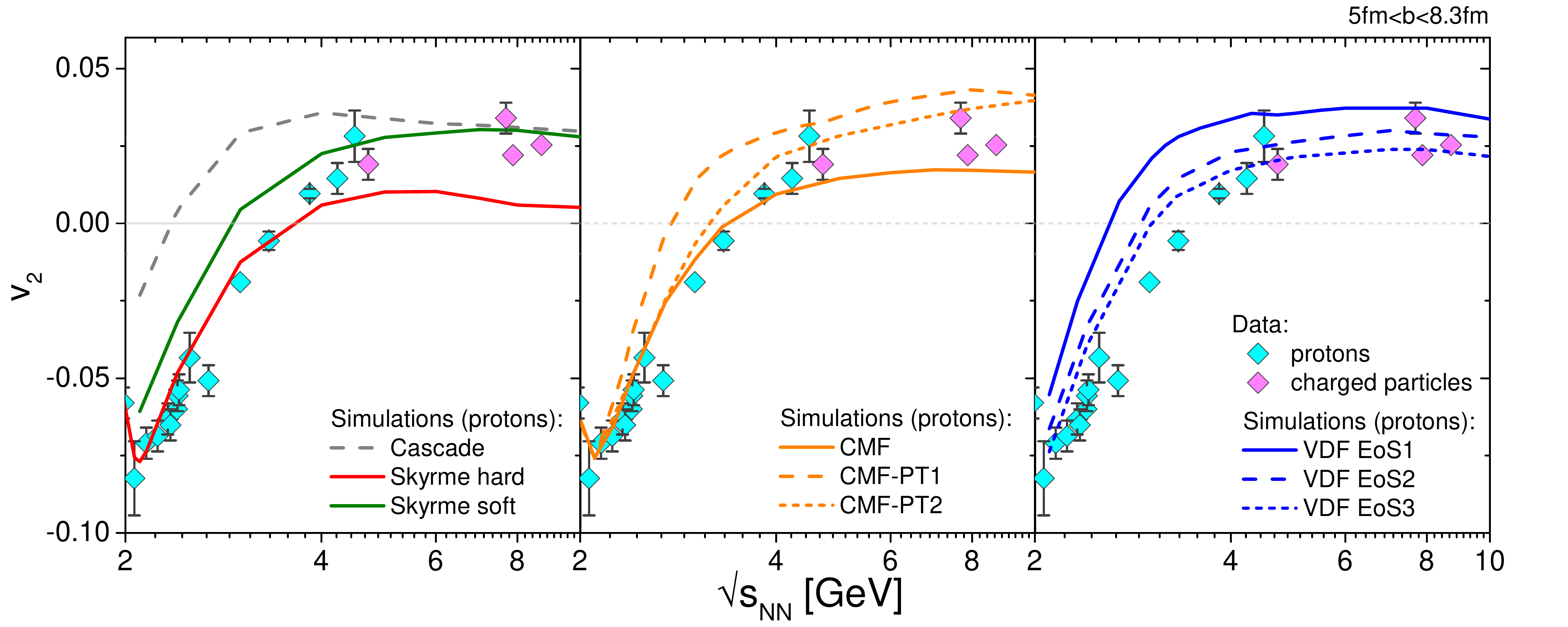}
  \caption{The elliptic flow of protons, with respect to the reaction plane, for mid-central Au--Au collisions ($5<b<8.3$ fm). The elliptic flow is evaluated at mid-rapidity $|y/y_b|<0.1 $ ($y$ is the proton rapidity in the center of mass frame of the collision, i.e. $y_\mathrm{cm} = 0$, and $y_b$ is the beam rapidity defined in the center-of-mass frame of the collision) and no cut in the transverse momentum is applied. From left to right, the results for the \texttt{UrQMD} cascade mode (grey dashed line), hard (red) and soft (green) Skyrme EoS, the CMF model (orange lines), and the VDF model (blue lines) are shown. All are compared to available data from heavy-ion collisions at several beam energies (symbols) \cite{E895:1999ldn,CERES:2002eru,FOPI:2004bfz,STAR:2012och,STAR:2020dav,HADES:2020lob,STAR:2021yiu}.
  }
  \label{fig:v2}
\end{figure*}

\section{Results on the flow at various beam energies}

The average radial flow is the most direct connection between the pressure gradient generated by the EoS and the final state observables. Therefore, we show the mean transverse kinetic energy of protons as obtained from our simulations, as a function of beam energy and compared to experimental data \footnote{Note that we chose to only present the mean transverse mass for the NA49 and STAR experiments, as the AGS data available often used different acceptances as well as centrality definitions, which makes a direct comparison difficult.}, in Fig.\ \ref{fig:mmt} for most central ($b<3.4$ fm) Au--Au collisions. The results for the mean transverse kinetic energy, as well as the other flow coefficients, are always evaluated in mid-rapidity defined as $|y/y_b|<0.1$, where $y$ is the proton rapidity in the center of mass frame of the collision, i.e. $y_\mathrm{cm} = 0$, and $y_b$ is the beam rapidity defined in the center-of-mass frame of the collision.
This is done to make the results comparable as the rapidity width of the colliding systems changes drastically in the beam energy range under investigation. 

The three panels in Fig.\ \ref{fig:mmt} present the different parametrizations of the density-dependent potential. The first panel shows the results for UrQMD in the cascade mode without any potentials (grey dashed line) compared to the well-known hard and soft Skyrme potentials (red and green solid lines). The UrQMD cascade mode resembles an ``ultra-soft'' EoS, and therefore the mean transverse flow is the smallest in this case. As expected, a harder EoS leads to a stronger transverse flow. In the comparison of the CMF (middle panel) and VDF (right panel) EoSs, the situation is more complicated. In general, the VDF model has a softer EoS for densities up to 2-3 times saturation density, which directly translates into a smaller transverse flow for beam energies between 3 and 5 GeV. The existence of a phase transition also leads to a softening and thus decreases the flow. Interestingly, the VDF EoS 1 and 2 show a rapid increase of the mean transverse flow at beam energies above $\sqrt{s_{\txt{NN}}} =6~ \txt{GeV}$ due to the strong hardening of the EoS for densities above 4 -- 5$n_0$ \footnote{We note that the speed of sound in the VDF model becomes superluminal for densities above the upper boundary of the coexistence region for the ``QGP-like'' phase transition. This is natural as the VDF EoS is not in any way constrained outside of the fitting region. Moreover, it can be shown that in any model using baryon number density to parametrize the interactions, $\lim_{n_B \to \infty} c_s^2 = b_{\txt{max}} - 1$, where $b_{\txt{max}}$ is the highest power of the baryon density entering the expression for the mean-field energy density in the system.}. It is clear that the density dependence of the EoS can be traced by the mean transverse momentum of protons. However, the available data is limited to the higher beam energies and additionally suffers from relatively large error bars. Comparing all EoSs used in our study, the standard CMF EoS as well as the VDF EoS3 seem to give the best description of available data due to their ``stiffness'' at low densities and subsequent softening at higher densities.

\begin{figure*}[t]
  \centering
  \includegraphics[width=\textwidth]{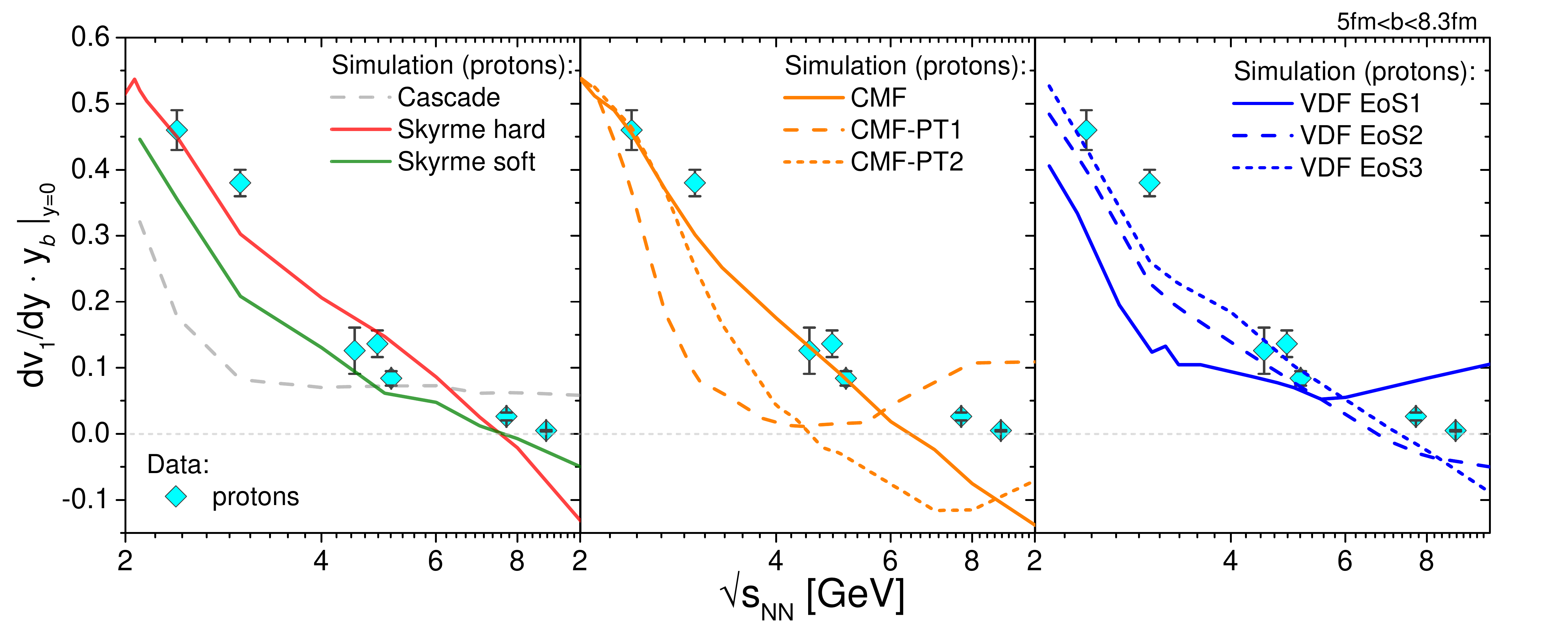}

  \caption{The slope of the directed flow, $d v_1/dy_{\mathrm{b}} \equiv d v_1/dy \cdot y_{b}$
($y$ is the proton rapidity in the center of mass frame of the collision, i.e. $y_\mathrm{cm} = 0$, and $y_b$ is the beam rapidity defined in the center-of-mass frame of the collision), for mid-central Au--Au collisions ($5<b<8.3$ fm). The slope is evaluated at mid-rapidity $|y/y_b|<0.1 $  and no cut in the transverse momentum is applied. From left to right, the results for the \texttt{UrQMD} cascade mode (grey dashed line), hard (red) and soft (green) Skyrme EoS, the CMF model (orange lines), and the VDF model (blue lines) are shown. All are compared to available data from heavy-ion collisions at several beam energies (symbols) \cite{STAR:2020dav,STAR:2021ozh,STAR:2021yiu,HADES:2020lob,HADES:2022osk,Kashirin:2020evw,E877:1996czs,NA49:2003njx}.
  }
  \label{fig:v1}
\end{figure*}

Higher order harmonics of the angular transverse flow distributions are also known to be rather sensitive to the EoS. One much investigated harmonic is the elliptic flow. The elliptic flow $v_2$ is mainly caused as a response of the medium to the asymmetric initial density distribution in non-central collisions. In this work we will calculate $v_2$ from:
\begin{equation}
    v_2 = \left\langle\frac{p_x^2 - p_y^2}{p_x^2 + p_y^2} \right\rangle ~,
\end{equation}
where the momenta are defined with respect to the reaction plane. 
The average runs over all protons in the mid-rapidity, $|y/y_b|<0.1$, region of the collision. As before, Fig.\ \ref{fig:v2} shows the results for $v_2$ for the different EoSs. Similar to the mean transverse flow, the elliptic flow from the UrQMD cascade mode shows the strongest deviation from the data due to its extreme ``softness''. In general, one can observe that the elliptic flow is increased when the EoS is softer \cite{Nara:2017qcg}. The beam energy dependence of $v_2$ also shows a clear EoS dependence: while at lower beam energies a harder EoS is favored, this is less true for energies above $\sqrt{s_{\mathrm{NN}}}> 3$ GeV. Here, a significant softening is possible. Overall, the models which have a moderate softening, either the CMF model with a crossover or the VDF models with a phase transition at higher density, give the best description of the $v_2$ at high beam energies. Similarly as in the case of the mean transverse flow, the elliptic flow appears to prefer an EoS which is stiff at moderate densities and shows a softening at higher densities.  

Finally, the directed flow and its dependence on the EoS is investigated. The directed flow $v_1$ and its derivative with respect to the rapidity has been discussed as a sensitive probe of the EoS \cite{Rischke:1995pe,Csernai:1999nf,Brachmann:1999xt,Stoecker:2004qu,Steinheimer:2014pfa,Nara:2016hbg}. It is defined as:
\begin{equation}
    v_1(y/y_b) = \left\langle\frac{p_x}{\sqrt{p_x^2 + p_y^2}} \right\rangle ~,
\end{equation}
where $y/y_b$ is the rapidity normalized by the beam rapidity at the center-of-mass frame, and the average runs over all protons in a given rapidity window. The derivative with respect to rapidity, often referred to as the slope of the directed flow, is calculated in the small bin around mid-rapidity $y/y_b=\pm 0.1$ to obtain the slope:
\begin{equation}
\left. dv_1/d(y/y_{b})\right|_{y=0} \equiv \left. d v_1/dy \cdot y_{b} \right|_{y=0}.  
\end{equation}
The results of our simulations are presented in Fig.
\ref{fig:v1} in a similar way as in the previous two figures \footnote{Again, we chose to not compare to directed flow observables from some of the AGS experiments as they presented a slightly different observable in different acceptance and centrality definitions, making a direct comparison difficult.}. Because the slope of the directed flow is a differential observable, its dependence on the EoS can be expected to be and indeed seems stronger than for the average transverse momentum $\left\langle m_T \right\rangle$ or elliptic flow $v_2$. Compared to the two Skyrme EoSs, the \texttt{UrQMD} cascade model gives the smallest slope of the directed flow, and the slope is increased for stiffer EoSs, at least for beam energies below $\sqrt{s_{\mathrm{NN}}}< 5$ GeV. A similar picture emerges for the CMF and VDF models. However, here the structure is more complex. In the scenarios where a strong phase transition occurs for densities below $4n_0$, a clear local minimum in the slope of $v_1$ can be observed; such behavior has been observed before in other fluid dynamic and transport simulations \cite{Stoecker:1980vf,Bozek:2010bi,Rischke:1995pe,Csernai:1999nf,Brachmann:1999xt,Steinheimer:2014pfa,Ohnishi:2017xjg}. If the phase transition occurs at a higher density, however, the minimum is not observed within the range of beam energies explored in this study. Because a local minimum can be considered a rather reliable observable, measurements of  the directed flow could be used to confirm or exclude certain classes of EoSs. Based on the currently available data, conclusions are difficult to draw, but with more precise data sets to become available in the future, $\left. dv_1/dy\right|_{y=0}$  can serve as a strong constraint on the EoS.

\section{Conclusions}

The role of a density-dependent EoS on different proton flow observables for collisions of heavy ions at the beam energy range of $\sqrt{s_{NN}}=2$ -- $10$ GeV was investigated. This constitutes the first time that an EoS with a phase transition was included in the molecular dynamics part of the microscopic transport model UrQMD. The method that was used to do so can be generalized for any density-dependent EoS.

It was found that all flow coefficients, $\left\langle m_T \right\rangle$, $v_1$, as well as $v_2$, are strongly influenced by the high density EoS up to approximately $4$ times saturation density. In particular, the effect of different first-order phase transitions were studied and it was confirmed that a strong transition would lead to a measurable decrease in the mean transverse mass, an increase in the elliptic flow and, at the same time, a minimum in the excitation function of the slope of the directed flow. In general, it was found that the currently available data favors a stiff equation of state up to $\sim 3-4$ times nuclear saturation density, after which a softening should occur. These results are qualitatively similar to the most likely EoS inferred from neutron star mergers and cold neutron stars which also hint at a maximum in the speed of sound \cite{Hebeler:2013nza,Bedaque:2014sqa,Fujimoto:2019hxv,Kojo:2021wax,Altiparmak:2022bke}. Also almost no sensitivity to the equation of state above 4 times saturation is observed. A more quantitative study on the available data will be performed in a future work.

\begin{acknowledgement}
A.M.\ acknowledges the Stern--Gerlach Postdoctoral fellowship of the Stiftung Polytechnische Gesellschaft.
J.S.\ thanks the Samson AG for funding. 
This work was supported by a PPP program of the DAAD.
A.S.\ acknowledges support by the U.S.\ Department of Energy, Office of Science, Office of Nuclear Physics, under Grant No.\ DE-FG02-00ER41132.
Y.N.\ acknowledges support by JSPS KAKENHI Grant Number JP21K03577.
V.K.\ acknowledges support by the U.S. Department of Energy, 
Office of Science, Office of Nuclear Physics, under contract number 
DE-AC02-05CH11231.
M.B.\ acknowledges support by the EU--STRONG 2020 network.
The computational resources for this project were provided by the Center for Scientific Computing of the GU Frankfurt and the Goethe--HLR.
\end{acknowledgement}

%\vspace{0.5 cm}

\appendix

\section{Parameter sets for the Skyrme model}
\label{s-para}

Parameter sets used for the soft and hard Skyrme parametrizations. The parameters enter in the single particle energy $U$ given in equation (\ref{eq:usky}).

\begin{table}[h]
\caption{Parameters used in the hard and soft Skyrme potential~\cite{Hartnack:1997ez}.\label{t1}}
\begin{tabular}{|c|c|c|}
\hline
Parameters & hard EoS & soft EoS \\
\hline
$\alpha$ [MeV] & -124 & -356 \\
\hline
$\beta$  [MeV] & 71 & 303 \\
\hline
$\gamma$ & 2.00 & 1.17\\
\hline
\end{tabular}

\end{table}

\section{Parameter sets for VDF1, VDF2, and VDF3 EoS}
\label{VDF_parameter_sets}

\begin{table*}[t]
	\caption{Parameter sets corresponding to the VDF EOSs reproducing nuclar matter characteristics listed in Table \ref{t1_VDF}.}
	\label{parameters}
	\begin{center}
		\bgroup
		\def\arraystretch{1.4}
		\begin{tabular}{c r r r r l l l l}
			\hline
			\hline
			%\cline{2-6}
			\hspace{2mm}set\hspace{2mm} & \hspace{4mm}$b_1$\hspace{4mm} &  \hspace{4mm} $b_2$  \hspace{4mm} &  \hspace{4mm}
			$b_3$  \hspace{4mm} &  \hspace{4mm} $b_4$  \hspace{4mm} &  \hspace{5mm} $\tilde{C}_1$ [MeV]  \hspace{2mm} &  \hspace{2mm} $\tilde{C}_2$ [MeV]  \hspace{2mm} &  \hspace{2mm}
			$\tilde{C}_3$ [MeV]  \hspace{2mm} &  \hspace{2mm}  $\tilde{C}_4$ [MeV]  \hspace{2mm}\\ 
			\hline
			VDF1 & 1.7681391   &  3.5293515    &   5.4352788  &    6.3809822   & \hspace{0mm}   -8.450948$\times 10^1$  &   3.843139$\times 10^1$ &  -7.958557$\times 10^0$   &   1.552594$\times 10^0$ \\
			VDF2 & 1.8025297   &   3.0777209   &   6.4303869   &  11.4003161  &\hspace{0mm} -9.1843484$\times 10^1$& 3.9574869$\times 10^{1}$  & -2.1547320$\times 10^{-1}$  &  4.5187616$\times 10^{-5}$ \\  
			VDF3 & 2.2138613  &    2.5261557    &  5.3081105    &  7.4901532 &\hspace{0mm} -3.130530$\times 10^2$& 2.611963$\times 10^{2}$  & -6.317680$\times 10^{-1}$  & 4.450564$\times 10^{-3}$ \\  
			\hline
			\hline
		\end{tabular}
		\egroup
	\end{center}
\end{table*}

In this appendix, we provide parameters corresponding to the EOSs reproducing sets of nuclear matter characteristics listed in Table \ref{t1_VDF}. The values of the coefficients of the interaction terms, $\{C_1, C_2, C_3, C_4 \}$, depend on a chosen system of units; here, we adopt a convention within which the single-particle potential is written as
\begin{eqnarray}
U = \sum_{k=1}^N  \tilde{C}_k \left( \frac{n_B}{n_0} \right)^{b_k - 1} ~,
\end{eqnarray}
where $n_0$ is the saturation density, so that $\tilde{C}_k$ must have a dimension of energy. Thus, $\tilde{C}_k$ and $C_k$ are related by
\begin{eqnarray}
C_k = \frac{\tilde{C}_k }{n_0^{b_k - 1}} ~. 
\end{eqnarray}

In Table \ref{parameters}, we list coefficients $\{\tilde{C}_1,  \tilde{C}_2, \tilde{C}_3, \tilde{C}_4\}$ in units of MeV. Note that in particular, the sum of all coefficients yields the (rest frame) value of the single-particle potential at $n_B = n_0$, $\sum_{i=1}^N \tilde{C}_i = -52.484 \ \txt{MeV}$.

\end{document}